\documentclass[%
 reprint,
showplace,showers,
showpacs,
showkeys,
nicotinuric,
 almsman,assume,
 API, perl
]{revtex4-1}

\usepackage{graphicx}
\usepackage{dcolumn}
\usepackage{bm}

\begin{document}

\newcommand{\beq} {\begin{equation}}
\newcommand{\enq} {\end{equation}}
\newcommand{\ber} {\begin {eqnarray}}
\newcommand{\enr} {\end {eqnarray}}
\newcommand{\eq} {equation}
\newcommand{\eqs} {equations }
\newcommand{\mn}  {{\mu \nu}}
\newcommand{\sn}  {{\sigma \nu}}
\newcommand{\rhm}  {{\rho \mu}}
\newcommand{\sr}  {{\sigma \rho}}
\newcommand{\bh}  {{\bar h}}
\newcommand {\er}[1] {equation (\ref{#1}) }
\newcommand {\ern}[1] {equation (\ref{#1})}
\newcommand{\mbf} {{ }}
\newcommand {\Er}[1] {Equation (\ref{#1}) }

\preprint{APS/NonbarchelPRL260416}

\title{A Conserved Cross Helicity for Non-Barotropic MHD}

\author{A. Yahalom}%
\email{asya@ariel.ac.il}
\affiliation{%
 Ariel University, Ariel 40700, Israel
}%

\date{\today}

\begin{abstract}

Cross helicity is not conserved in non-barotropic magnetohydrodynamics (MHD)
(as opposed to barotropic or incompressible MHD). Here we show that
variational analysis suggests a new kind of cross helicity which is conserved in the non
barotropic case. The non barotropic cross helicity reduces to the standard cross helicity under barotropic
assumptions. The new cross helicity is conserved even for topologies for which the variational principle does
not apply.

\end{abstract}



\maketitle

Cross Helicity was first described by Woltjer \cite{Woltjer1,Woltjer2} and is give by:
\begin{equation} \label{GrindEQ__22_}
H_{C} \equiv \int  \vec{B}\cdot \vec{v}d^{3} x,
\end{equation}
in which $\vec{B}$ is the magnetic field, $\vec{v}$ is the velocity field and the integral is taken
over the entire flow domain. $H_{C}$ is conserved for barotropic or incompressible MHD and
is given a topological interpretation in terms of the knottiness of magnetic and flow field lines.
An analogous conserved helicity for fluid dynamics was obtained by Moffatt \cite{Moffatt}.
Both conservation laws for the helicity in the fluid dynamics case and the barotropic MHD case were
shown to originate from a relabelling symmetry through the Noether theorem \cite{Yahalomhel,Padhye1,Padhye2,YaLy}.

Consider the equations of non-barotropic MHD \cite{Sturrock,Yahalom1}:
\beq
\frac{\partial{\vec B}}{\partial t} = \vec \nabla \times (\vec v \times \vec B),
\label{Beq}
\enq
\beq
\vec \nabla \cdot \vec B =0,
\label{Bcon}
\enq
\beq
\frac{\partial{\rho}}{\partial t} + \vec \nabla \cdot (\rho \vec v ) = 0,
\label{masscon}
\enq
\beq
\rho \frac{d \vec v}{d t}=
\rho (\frac{\partial \vec v}{\partial t}+(\vec v \cdot \vec \nabla)\vec v)  = -\vec \nabla p (\rho,s) +
\frac{(\vec \nabla \times \vec B) \times \vec B}{4 \pi},
\label{Euler}
\enq
\beq
 \frac{d S}{d t}=0.
\label{Ent}
\enq
In the above the following notations are utilized: $\frac{\partial}{\partial t}$ is the temporal derivative,
$\frac{d}{d t}$ is the temporal material derivative and $\vec \nabla$ has its
standard meaning in vector calculus. $\rho$ is the fluid density and $S$ is the specific entropy. Finally $p (\rho,S)$ is the pressure which
depends on the density and entropy (the non-barotropic case).  \Er{Beq}describes the
fact that the magnetic field lines are moving with the fluid elements ("frozen" magnetic field lines),
 \ern{Bcon} describes the fact that the magnetic field is solenoidal, \ern{masscon} describes the conservation of mass and \ern{Euler}
is the Euler equation for a fluid in which both pressure
and Lorentz magnetic forces apply. \Er{Ent} describes the fact that heat is not created (zero viscosity, zero resistivity)
 in ideal non-barotropic MHD and is not conducted, thus only convection occurs.
The number of independent variables for which one needs to solve is eight
($\vec v,\vec B,\rho,S$) and the number of \eqs (\ref{Beq},\ref{masscon},\ref{Euler},\ref{Ent}) is also eight.
Notice that \ern{Bcon} is a condition on the initial $\vec B$ field and is satisfied automatically for
any other time due to \ern{Beq}.

In non-barotropic MHD one can calculate the temporal derivative of the cross helicity (\ref{GrindEQ__22_}) using the
above equations and obtain:
\begin{equation} \label{GrindEQ__22c_}
\frac{dH_{C}}{dt}= \int T  \vec{\nabla} S \cdot \vec{B} d^{3} x,
\end{equation}
in which $T$ is the temperature. Hence, generally speaking cross helicity is not conserved.

A clue on how to define cross helicity for non\-barotropic MHD can be obtained from
the variational analysis described in \cite{Yahalom1} which is valid for magnetic
field lines at the intersection of two comoving surfaces $\chi,\eta$ (Euler potentials).
Following Sakurai \cite{12} the magnetic field takes the form:
\begin{equation} \label{GrindEQ__10_}
\vec{B}=\vec{\nabla }\chi \times \vec{\nabla }\eta .
\end{equation}
In terms of the functions $\chi_i\equiv (\chi,\eta,S)$ one obtains the five function Lagrangian density \cite{Yahalom1}:
\ber
& & \hat {\cal L}[\chi_i,\nu,\rho] = \rho [\frac{1}{2} A^{-1}_{jn} \frac{\partial \chi_j}{\partial t}  \frac{\partial \chi_n}{\partial t}
+\frac{\partial \nu}{\partial \chi_m} \frac{\partial{\chi_m}}{\partial t}
 \nonumber \\
&-&  \frac{\partial{\nu}}{\partial t} -\  \varepsilon (\rho,\chi_3)]
- \frac{1}{8 \pi}(\vec \nabla \chi_1 \times \vec \nabla \chi_2)^2.
\label{Lagactionsimp8}
\enr
The two additional functions on which the Lagrangian density depend are a Bernoulli type function $\nu$ and the mass
density $\rho$. The Lagrangian density depend on the $\chi_i$ fields through the inverse of the symmetric $A$ matrix
defined as:
\beq
A_{ij} \equiv  \vec \nabla \chi_i \cdot \vec \nabla \chi_j
\label{Adef}
\enq
and through the specific internal energy $\varepsilon$ which is a function of density and entropy.
Einstein summation convention is assumed throughout.

Variational principles for magnetohydrodynamics were introduced by
previous authors both in Lagrangian and Eulerian form. Sturrock
\cite{Sturrock} has discussed in his book a Lagrangian variational
formalism for magnetohydrodynamics.
Vladimirov and Moffatt \cite{VMoffatt} in a series of papers have discussed an Eulerian
variational principle for incompressible magnetohydrodynamics.
However, their variational principle contained three more
functions in addition to the seven variables which appear in the
standard equations of incompressible magnetohydrodynamics which are the magnetic
field $\vec B$ the velocity field $\vec v$ and the pressure $P$.
Kats \cite{Kats} has generalized Moffatt's work for compressible
non barotropic flows but without reducing the number of functions
and the computational load.  Sakurai \cite{12} has introduced a two function Eulerian variational
principle for force-free magnetohydrodynamics and used it as a
basis of a numerical scheme, his method is discussed in a book by
Sturrock \cite{Sturrock}. Yahalom \& Lynden-Bell \cite{YaLy} combined the Lagrangian of
Sturrock \cite{Sturrock} with the Lagrangian of Sakurai
\cite{12} to obtain an {\bf Eulerian} Lagrangian principle for barotropic magnetohydrodynamics
which will depend on only six functions. The variational
derivative of this Lagrangian produced all the equations
needed to describe barotropic magnetohydrodynamics without any
additional constraints. The equations obtained resembled the
equations of Frenkel, Levich \& Stilman \cite{FLS} (see also \cite{Zakharov}).
Yahalom \cite{Yah} have shown that for the barotropic case four functions will
suffice. Moreover, it was shown that the cuts of some of those functions \cite{Yah2}
are topological local conserved quantities.

Previous work was concerned only with
barotropic magnetohydrodynamics. Variational principles of non
barotropic magnetohydrodynamics can be found in the work of
Bekenstein \& Oron \cite{Bekenstien} in terms of 15 functions and
V.A. Kats \cite{Kats} in terms of 20 functions.
Morrison \citep{Morrison} has suggested a Hamiltonian approach but this also depends on 8 canonical variables (see table 2 \citep{Morrison}).
The variational principle introduced  in \cite{Yahalom1} show that only five functions will suffice to
describe non barotropic MHD in the case
that we enforce a Sakurai \cite{12} representation for the magnetic field.

The variational equations are given in terms of the
quantities:
\beq
\alpha_i \equiv (\alpha, \beta, \sigma), \quad
 \alpha_i [\chi_i,\nu]= - A^{-1}_{ij} (\frac{\partial \chi_j}{\partial t} + \vec \nabla \nu \cdot \vec \nabla \chi_j).
 \label{ali}
\enq
And the generalized Clebsch representation of the velocity \cite{Yahalom1}:
\beq
\vec v =  \vec \nabla \nu + \alpha \vec \nabla \chi + \beta \vec \nabla \eta + \sigma \vec \nabla S.
\label{vform}
\enq
as follows:
\ber
\frac{d \nu}{d t} &=& \frac{1}{2} \vec v^2 - w,
\label{eqnu} \\
\frac{\partial{\rho}}{\partial t} &+& \vec \nabla \cdot (\rho \vec v ) = 0,
\label{eqrho} \\
\frac{d \sigma}{dt} &=& T,
\label{eqsigma} \\
\frac{d \alpha}{dt} &=& \frac{\vec \nabla \eta \cdot \vec J}{\rho},
\label{eqalpha} \\
\frac{d \beta}{dt} &=& -\frac{\vec \nabla \chi \cdot \vec J}{\rho}.
\label{eqbeta}
\enr
In the above: $w$ is the specific enthalpy and the current is $\vec J =\frac{\vec \nabla \times \vec B}{4 \pi}$.
The above equations are shown \cite{Yahalom1} to be equivalent to
the non barotropic MHD \eqs (\ref{Beq}-\ref{Ent}).

The function $\nu$ whose material derivative is given in \ern{eqnu}
can be multiple valued as only its gradient appears in the velocity (\ref{vform}). However,
the discontinuity of $\nu $ is a conserved quantity :
\begin{equation} \label{GrindEQ__37_}
\frac{d[\nu ]}{dt} =0.
\end{equation}
since the right hand side of \ern{eqnu} are physical and hence single valued quantities.
A similar equation hold also for barotropic fluid dynamics and barotropic MHD \cite{YaLy2,YaLy,Yah,Yah2}.

Let us now write the cross helicity given in \ern{GrindEQ__22_} in terms of \ern{GrindEQ__10_} and \ern{vform},
this will take the form:
\begin{equation} \label{GrindEQ__22d_}
H_{C} = \int d \Phi [\nu] + \int d \Phi \oint \sigma d S
\end{equation}
in which: $d\Phi =\vec{B}\cdot d\vec{A}=\vec{\nabla }\chi \times \vec{\nabla }\eta \cdot d\vec{A}=d\chi \, d\eta $ and the closed line
integral is taken along a magnetic field line. $d\Phi$ is a magnetic flux element which is
comoving according to \ern{Beq} and $d\vec{A}$ is an infinitesimal area element. Although the
cross helicity is not conserved for non-barotropic flows, looking at the right hand side we see
that it is made of a sum of two terms. One which is conserved as both $d\Phi$ and $[\nu]$ are comoving
and one which is not. This suggests the following definition for the non barotropic cross helicity $H_{CNB}$:
\begin{equation} \label{GrindEQ__22e_}
H_{CNB} \equiv \int d \Phi [\nu] = H_{C} -  \int d \Phi \oint.
\sigma d S
\end{equation}
Which can be written in a more conventional form:
\begin{equation} \label{GrindEQ__22e1_}
H_{CNB} = \int  \vec{B} \cdot \vec v_t  d^{3} x
\end{equation}
in which the topological velocity field is defined as follows:
\begin{equation} \label{vt_}
 \vec v_t = \vec v - \sigma \vec \nabla S
\end{equation}
It should be noticed that $H_{CNB}$ is conserved even for an MHD not satisfying
the Sakurai topological constraint given in \ern{GrindEQ__10_}, provided that we have
a field $\sigma$ satisfying the equation $\frac{d \sigma}{dt} = T$. Thus the non barotropic cross helicity
conservation law:
\begin{equation} \label{HCNBcon}
\frac{d H_{CNB}}{dt} = 0,
\end{equation}
is more general than the variational principle described by \ern{Lagactionsimp8} as follows from a
direct computation using  \eqs (\ref{Beq},\ref{masscon},\ref{Euler},\ref{Ent}). Also notice that for a constant
specific entropy $S$ we obtain $H_{CNB}=H_{C}$ and the non-barotropic cross helicity reduces to the standard barotropic cross helicity.
To conclude we introduce also a local topological conservation law in the spirit of \cite{Yah2} which is the
non barotropic cross helicity per unit of magnetic flux. This quantity which is equal to the discontinuity of $\nu$  is conserved and
 can be written as a sum of the barotropic cross helicity per unit flux and the closed line integral of $S d \sigma$ along a magnetic field line:
\begin{equation} \label{loc_}
[\nu]= \frac{dH_{CNB}}{d \Phi} = \frac{dH_{C}}{d \Phi} + \oint  S d \sigma.
\end{equation}

\end {document}